# Examining the Influence of Job Satisfaction on Individual Innovation and Its Components: Considering the Moderating Role of Technostress

**Fatemeh Daneshmandi,** Academic Center for Education, Culture, and Research, Neyshaboor, Iran s.daneshmandi69@gmail.com

**Hassan Hessari**, Department of Business Information, Technology, Pamplin College of Business, Virginia Tech, Blacksburg, Virginia, USA hassanhessari@vt.edu

**Tahmineh Nategh**, Department of Management, Shahrood Branch, Islamic Azad University, Shahrood, Iran tnategh@iau-shahrood.ac.ir

**Ali Bai,** College of Business, University of Northern Iowa, Cedar Falls, Iowa, USA ali.bai@alumni.uni.edu

## Abstract

*Background:* Employee innovation is a crucial aspect for organizations in the current era. Therefore, studying the factors influencing individual innovation is vital and unavoidable. Undoubtedly, job satisfaction is a significant variable in management sciences. Nowadays, all organizations are interconnected with technology.

*Objective:* This research explores the relationship between job satisfaction and individual innovation, including its components, and the moderating role of technostress.

*Research Method:* This study, in terms of purpose, is applied, and in terms of data collection method, it is descriptive-survey. Data collection tools included the Technostress Inventory by Tarafdar and colleagues (2007), Janssen's Individual Innovation Questionnaire (2000), and the Job Satisfaction Survey (JSS) by Spector (1994). The validity and reliability of these questionnaires were confirmed. The sample size for this study was 215, and data analysis was performed using SPSS and SMART-PLS software.

*Findings:* Job satisfaction has a significant and positive relationship with individual innovation, idea generation, idea promotion, and idea implementation. Technostress moderates the relationship between job satisfaction and individual innovation, as well as idea generation and idea promotion. However, technostress does not play a moderating role in the relationship between job satisfaction and idea implementation.

*Conclusion:* Based on the obtained results, organizations should take necessary measures to increase job satisfaction and reduce technostress among their employees.

**Keywords:** Technostress, Job Satisfaction, Individual Innovation, Components of Individual Innovation, Organization

# 1. Introduction

In recent years, it has become common among researchers and scholars that innovation is crucial for organizations (Drach-Zahavy & et al., 2004). Countries around the world, relying on innovation, aim to increase productivity and improve their economic conditions. One of the major reasons for this focus is the increasing competition among societies (Mirfakhredini et al., 2010). Moreover, the rise in complexities and environmental uncertainties necessitates attention to innovation. Organizations that excel are those with high adaptability and the ability to enhance their service quality (Robbins, 2007). Today, there is a consensus on the significance of innovation in organizational context and the need for markets to be more dynamic. Organizations must innovate to meet the changing demands and lifestyles of their customers and to exploit opportunities created by technology and market changes (Rowley & et al., 2011).

According to Chen et al. (2004), innovation refers to the introduction of a new combination of essential elements in the production system. Innovation capital includes organizational competency, research and development execution, and the creation of new technology and products to meet customer demands. The Organization for Economic Cooperation and Development (OECD) refers to any commercial exploitation of new knowledge as innovation. In other words, innovation entails implementing new processes or procedures that fundamentally differ from what already exists (Du Plessis, 2007).

In this environment, organizations must ensure that their strategies are innovative and new to maintain their competitive advantage. Successful innovative activities within an organization require employees' participation at all levels (Lloréns Montes et al., 2004). Employees are considered crucial for fostering innovation. While innovation in literature is often rooted in creativity, such innovations are relatively rare, whereas gradual innovations based on employees' efforts are more common. Undoubtedly, a significant aspect for any organization is the discussion of individual innovation among its employees. The key to any organization's success and survival today lies in the innovativeness of its employees and encouraging them to innovate (Osayawe Ehigie & Clement Akpan, 2004). Individual innovations are central to many principles in modern management, including Total Quality Management (Osayawe Ehigie & Clement Akpan, 2004), Continuous Improvement initiatives (Fuller & et al., 2006), corporate audacity (Elfring, 2003), creative problem-solving (Basadur, 2004), and organizational learning (Senge, 1990).

Today, individual innovation is recognized as a vital element for effective organizational performance and long-term sustainability (Janssen, 2000). Innovation behavior refers to individuals' deliberate efforts to generate, introduce, and utilize new ideas. Researchers generally agree that individual innovation starts with problem recognition and idea generation but also includes advocating for ideas and trying to build coalitions of supporters for idea implementation (Janssen, 2000; Kanter, 2009; Scott & Bruce, 1994).

Previous studies indicate that organizations can indeed benefit from individual innovation (Campbell et al., 1996). Empirically, a positive relationship between specific innovation behaviors and organizational performance has been demonstrated (Miron et al., 2004). It has also been shown that individual innovation does not diminish the quality and efficiency of normal work. Employees are well capable of maintaining a balance between the presence of innovation and the organizational features necessary for innovation promotion; quality and efficiency complement each other rather than compete. In this regard, Getz & Robinson (2003) proposed an interesting rule of thumb: companies that improved through idea resources found that 80% of these ideas were improvements suggested by their own employees, and only 20% were planned innovation activities. This validates the organization's capacity to harness new products and other aspects of performance through the knowledge of its human resources (Foss, 2007). Therefore, the most unique and non-replicable resources available to companies are individuals with knowledge, who can effectively utilize other organizational resources (Argote & Ingram, 2000).

Therefore, paying attention to employees' conditions is vital for any organization. One of the influential factors on employees' conditions and, consequently, on the organization, is job satisfaction. The concept of job satisfaction among employees is so critical in the literature of human relations and organizational behavior that researchers have examined various predictors of this key element in organizational achievement (Kifle & et al., 2012). Not having job satisfaction among employees is one of the major problems for organizations, leading to employee turnover, absenteeism, organizational disengagement, low motivation, and ultimately, employees leaving the organization (Nasrollahi et al., 2022). This issue comes with high costs for the organization, including decreased productivity, training costs, empowerment, recruitment, and loss of organizational knowledge (Ansari et al., 2010).

On the other hand, job satisfaction is one of the influential variables in organizational efficiency, the high level of which can lead to increased organizational dynamism and productivity (Hessari & Nategh, 2022a; Kwai & et al., 2010). Job satisfaction is a positive emotional state or feeling resulting from job evaluation or individual experience (Sepahvand et al., 2023). This positive emotional state significantly contributes to individuals' physical and mental well-being. From an organizational perspective, a high level of job satisfaction reflects a highly desirable organizational atmosphere and leads to the attraction and retention of employees (Mohaghar et al., 2022).

Many researchers have attempted to define job satisfaction. For instance, Hopkins defined job satisfaction as "meeting specific needs related to personal work" and its spectrum as "liking the job." Also, job satisfaction has been defined as a positive orientation of an individual toward the job role they currently perform (Griffin & et al., 2010) and a pleasurable mental state because of evaluating one's job as facilitating or meeting one's self-values. In general, satisfaction and dissatisfaction are evaluated as a function of the observed relationship between what an individual wants from their job and what they get; hence, job satisfaction is an attitude indicating an individual's satisfaction, contentment, or persuasion in their work or job (Sharma & et al., 2010). Studies indicate that since 1976, over 3000 studies related to job satisfaction have been conducted. Researchers have examined the impact of various factors and variables on job satisfaction. Influential factors on job satisfaction include social health, mental health, leadership style, salary, promotion process, working conditions, and the job itself (Hakkak et al., 2022; Griffin & et al., 2010). In summary, job satisfaction holds significance in three dimensions:

- *Individual Dimension:* This aspect determines its influence on employees' behavior in performing their tasks efficiently.
- *Organizational Dimension:* This refers to the impact of job satisfaction on organizational variables. Managers should prioritize their employees' job satisfaction for three reasons: Dissatisfied individuals are more likely to be absent, resign, or leave the work environment. Satisfied employees experience better physical and mental health and tend to live longer. Increased job satisfaction leads to higher commitment to work (organizational commitment).
- *Social Dimension:* This signifies the impact of job satisfaction on "overall life satisfaction".

Additionally, it is undeniable that technology has become an inseparable part of today's society to the extent that organizations or individuals not engaging with technology are rarely found. Technology has a direct impact on organizational survival; hence, analyzing the effects of technology on organizations is crucial and necessary (Hessari et al., 2023). Successful organizations have always endeavored to utilize technology appropriately to advance their goals and ensure organizational sustainability. Positive effects of the technological revolution in organizations include improving productivity and reducing workplace fatigue (Vieitez & et al., 2001). However, alongside these positive effects, technology also has negative impacts on organizations. In today's rapidly changing technological world, technology introduces destructive stress known as "technostress" into organizations (Rouhani & Mohammadi, 2022). This has led many employees to suffer from technostress. Previous literature indicates that the technological revolution contributes to an increase in job-related stress in the workplace (Rosen & Weil, 2000; Gallie, 2005).

Brod, a consultant, and psychologist specializing in adapting to new technology, introduced technostress in his book titled "Technostress: The Human Cost of the Computer Revolution" as a modern disease arising from humans' inability to adapt to new computer technologies in a healthy manner (Brod, 1984). Although Brod viewed technostress as an illness, other researchers consider it as an inability to adapt to changes imposed by technology. This reduced productivity and effectiveness in individuals' tasks. This decrease in productivity lowers the organization's success rate. Technology, especially information technology, is rapidly evolving, and consequently, organizations that do not update themselves lose their competitive edge against other organizations (Hessari & Nategh, 2022b). The implementation of new technologies can create stress among organizational employees and negatively impact productivity. While IT might be costly for an organization, it is essential for accessing timely and relevant information to make appropriate decisions (Tarafdar & et al., 2007; Weil & Rosen, 1997). Tarafdar and colleagues developed a measurement scale for technostress based on the data from the United States. They identified five elements of techno-stress, which are recognized as techno-stress builders (see Table 1).

Table 1: Techno-Stress Components (Tarafdar et al., 2007).

| Technostress Components | Description |
| --- | --- |
| Technology Overload | A situation where an individual is compelled to work faster and longer due to technology. |
| Invasion of Technology | A situation in which employees feel technology can keep them constantly connected to work, due to the blurred boundary between work and personal matters. |
| Techno-Complexity | A situation where employees feel their skills are insufficient to handle the complexities related to information and communication technologies. Consequently, they are forced to invest time and effort in learning and understanding various aspects of technology. |
| Techno-Insecurity | A situation where employees feel threatened that they might lose their jobs to new information and communication technologies, or they might be replaced by others who are better at using these technologies. |
| Technology Uncertainty | A situation where employees feel uncertain and restless due to the rapidly changing nature of information and communication technologies, requiring constant updating. |

The healthcare industry has also witnessed an increase in technostress. With the proliferation of online journals and research articles, medical facilities are inundated with vast amounts of information. While this information can be useful, it also leads to the emergence of technostress. Researchers have learned that by using technology, they can filter and consolidate necessary information in a targeted manner to reduce information overload. However, even with this filtration, a considerable amount of information remains, leaving users anxious as they are unable to compare all of it (Hall & Walton, 2004).

Various research results clearly indicate that employees can become tired and disheartened from technology (Moore, 2000; Murray & Rostis, 2007; Sethi & Barrier, 1999). Techno-fatigue causes employees to lose their efficiency. Managing technostress can be challenging for an organization. Weil and Rosen found that scientific evidence shows technostress also leads to excessive work perception, excessive information, loss of motivation, and job dissatisfaction (Weil & Rosen, 1997).

Brenda Mack and colleagues (2010) in their article titled "Technostress and Organizational Loyalty in IS&T Employees" investigated this matter. Their results indicate that understanding job stress reduces job satisfaction, while a better technology management strategy creates job satisfaction and organizational commitment (Mack & et al., 2010). In line with this, Rajesh Kamur and colleagues

(2013) in their article titled "The Relationship of Technostress on Job Satisfaction and Organizational Commitment among Professional (IT) Employees" explored this topic. Their study, conducted with 80 IT professionals from a technology park, clearly showed that technostress has a negative correlation with organizational commitment and job satisfaction (Kumar & et al., 2013).

Gaiter and colleagues (2008) used structural equation modelling to examine the relationship between organizational environment, work-family conflict, job stress, individual characteristics, job satisfaction, and organizational commitment in pharmacists in the United States. The results showed that organizational factors such as excessive workload, conflicts, and easy efficiency were among the most important variables that had the greatest impact on job stress. Moreover, their model revealed a significant relationship between individual characteristics, job satisfaction, and organizational commitment (Gaiter & et al., 2008). Building upon the preceding statement, we formulate the following hypotheses:

**H1:** Job satisfaction has a significant and positive impact on individual innovation.
**H2:** Technostress moderates the relationship between job satisfaction and individual innovation.
**H3:** Job satisfaction has a significant and positive impact on idea generation component.
**H4:** Job satisfaction has a significant and positive impact on idea promotion component.
**H5:** Job satisfaction has a significant and positive impact on idea implementation component.
**H6:** Technostress moderates the relationship between job satisfaction and idea generation component.
**H7:** Technostress moderates the relationship between job satisfaction and idea promotion component.
**H8:** Technostress moderates the relationship between job satisfaction and idea implementation component.

## 2. Research Methodology

This research is categorized as applied research in terms of its objectives and is a descriptive-survey study using correlational analysis through structural equation modelling in data collection. Data for this study were obtained through standardized questionnaires distributed in person. This research was conducted at several Organizations in Mashhad, Iran.

The researcher distributed 260 questionnaires randomly, following Krejcie & Morgan's table (1970). Out of these, 215 questionnaires were valid and usable. The data were collected in the first half of the year 2021. The Job Satisfaction Survey (JSS) questionnaire (1994), consisting of 4 questions on satisfaction with pay, 4 questions on job promotion, 4 questions on supervision, 4 questions on fringe benefits, 4 questions on potential rewards, 4 questions on job execution processes, 4 questions on co-workers, 4 questions on the nature of the job, and 4 questions on communication, was used. Additionally, the Techno-Stress questionnaire by Tarafdar et al. (2007), consisting of 6 questions on technology overload, 3 questions on technology invasion, 5 questions on technology complexity, 5 questions on technology insecurity, and 4 questions on technology uncertainty, was used. The Individual Innovation questionnaire by Jansen (2000), consisting of 3 questions on idea generation,

3 questions on idea promotion, and 3 questions on idea implementation, was used. The Likert scale, ranging from "completely disagree" as the starting point to "completely agree" as the endpoint, was used for measurement. The scores for the questions were calculated from 1 to 5.

After data collection, the obtained data from the questionnaires were transferred to the raw data pages of SPSS and SMART-PLS software for analysis. Descriptive statistics, including frequency and percentage, and inferential statistics, including structural equation modelling, were employed for analysis. The structural equation modelling analysis (path analysis) was conducted using SMART-PLS software.

To assess the validity of the measurement tools used in this study, feedback from university professors and experts was gathered. Additionally, since the items included in these questionnaires were designed based on standard questionnaires, the mentioned questionnaires have good validity. To assess reliability, an initial sample of 30 questionnaires was pre-tested, and then, using the data obtained from these questionnaires and with the help of SPSS statistical software, the coefficient of confidence was calculated using the Cronbach's alpha method. The Cronbach's alpha coefficient was calculated as 0.962 for the Technostress questionnaire, 0.982 for the Job Satisfaction questionnaire, and 0.962 for the Individual Innovation questionnaire. Moreover, the Cronbach's alpha coefficients for the components of Individual Innovation (idea generation 0.940, idea promotion 0.957, idea implementation 0.936) were calculated. Therefore, it can be said that the questionnaires used in this study have high internal consistency, reliability, and validity.

## 3. Data Analysis and Findings

In this study, structural equation modelling was employed to analyze and evaluate the model. Structural equation modelling is a statistical technique used to examine linear relationships between observed (measured) variables and latent (unobserved) variables. In other words, structural equation modelling is a powerful statistical technique that combines measurement modelling (confirmatory factor analysis) and structural modelling (regression or path analysis) with a simultaneous statistical test. For this analysis, SMART-PLS software was used. This software analyzes structural equation models containing several variables and encompasses direct, indirect, and interaction effects. Furthermore, the outputs from this analysis are presented and discussed below.

The frequency and percentage of the research sample group based on gender, education level, and work experience are presented in Table 2. Considering the outputs, the model's factor loadings on the model (Figure 1) and the statistical significance (T statistics) of the model's relationships (Figure 2), the findings of the research indicate that job satisfaction has a strong, positive, and significant impact on individual innovation. On the other hand, technostress moderates the relationship between job satisfaction and individual innovation. The study further examines the impact on the components of individual innovation.

The findings demonstrate that job satisfaction significantly and positively influences the components of idea generation, idea promotion, and idea implementation. Additionally, technostress moderates the relationship between job satisfaction and the components of idea generation and idea promotion. However, technostress does not have a moderating role in the relationship between job satisfaction and the component of idea implementation. The exact path coefficients and T statistics are presented in Table 3.

Table 2 - Frequency and Percentage of the Research Sample Group.

| Variable | Subgroups | Frequency |
|---|---|---|
| Gender | Male | 125 |
|  | Female | 90 |
| Age | 22 to 30 | 69 |
|  | 31 to 39 | 76 |
|  | 40 to 48 | 48 |
|  | 49 to 57 | 22 |
| Education Level | Diploma | 19 |
|  | Bachelor's Degree | 92 |
|  | Master's Degree | 86 |
|  | Doctorate | 18 |
| Work Experience | 1 to 4 years | 45 |
|  | 5 to 9 years | 66 |
|  | 10 to 14 years | 57 |
|  | 15 to 19 years | 30 |
|  | 20 to 40 years | 17 |

Figure 1 - Factor Loadings on the Path.

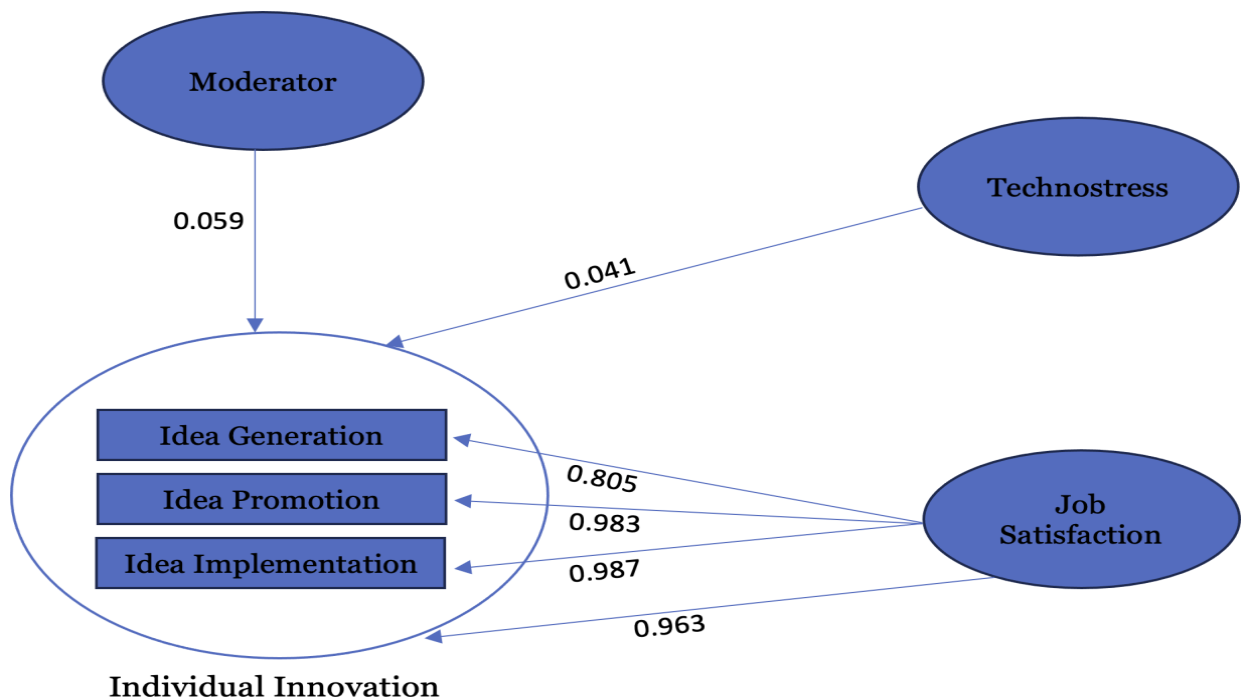

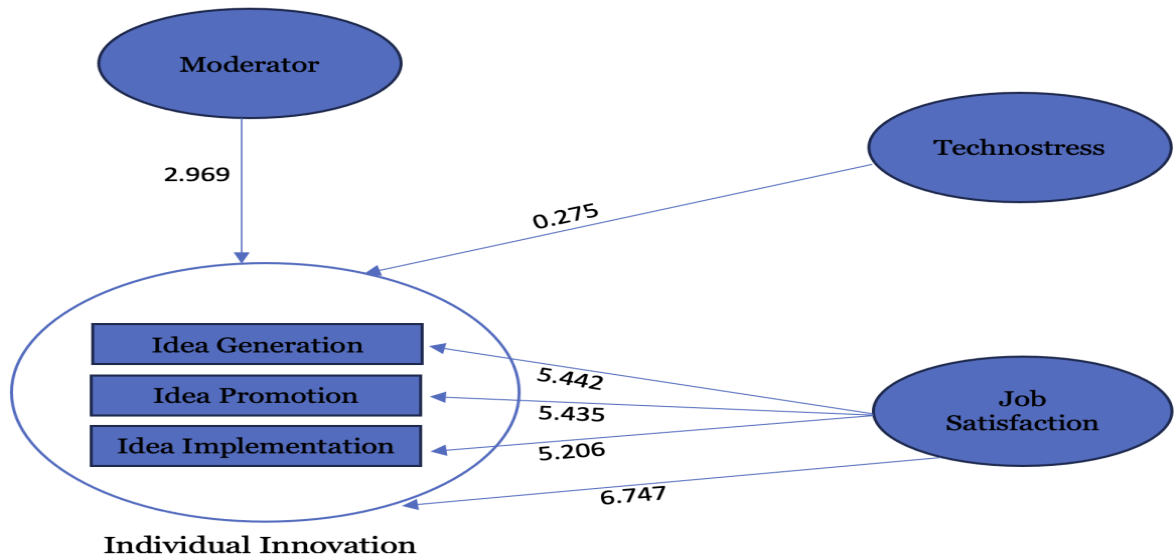

Figure 2 - T-Statistic Values.

Table 3 - Hypotheses Examination, Path Coefficients, and T-Statistics.

| Hypothesis | Path | Loading Factors | T-Statistic | Confirmation Rejection |
|---|---|---|---|---|
| 1 | Job Satisfaction → Individual Innovation | 0.963 | 6.747 | Confirmed |
| 2 | Technostress Moderating the Relationship between Job Satisfaction and Individual Innovation | 0.059 | 2.969 | Confirmed |
| 3 | Job Satisfaction → Idea Generation Component | 0.805 | 5.442 | Confirmed |
| 4 | Job Satisfaction → Idea Promotion Component | 0.983 | 5.435 | Confirmed |
| 5 | Job Satisfaction → Idea Implementation Component | 0.987 | 5.206 | Confirmed |
| 6 | Technostress Moderating the Relationship between Job Satisfaction and Idea Generation Component | 0.079 | 3.216 | Confirmed |
| 7 | Technostress Moderating the Relationship between Job Satisfaction and Idea Promotion Component | 0.082 | 2.708 | Confirmed |
| 8 | Technostress Moderating the Relationship between Job Satisfaction and Idea Implementation Component | 0.011 | 0.292 | Rejected |

Finally, considering the research findings, it should be noted that in Partial Least Squares (PLS) models, two types of models are examined. The Outer Model, equivalent to the measurement model, and the Inner Model, like the structural model in other software models such as LISREL, EQS, and AMOS. For evaluating the fit of the Outer Model, the Average Communality was used, and for the structural model fit, R2 was employed. The Average Communality indicates the percentage of indicator variance explained by the corresponding constructs and researchers have suggested an acceptable level for communality greater than 0.5 (Lee & et al,. 2008). The R2, which demonstrates the model's ability to explain the structure, is higher than 0.5, indicating a suitable fit for the proposed model, as presented in Table 4.

Table 4 - Model Fit.

| Variables | Average Communality | R2 |
|---|---|---|
| Job Satisfaction | 0.8879 | - |
| Individual Innovation | 0.93 | 0.8718 |
| Technostress | 0.8703 | - |
| Idea Generation | 0.9343 | 0.7969 |
| Idea Promotion | 0.9151 | 0.8091 |
| Idea Implementation | 0.9406 | 0.8292 |
| Technostress × Job Satisfaction | 0.9594 | - |

## 4. Discussion

This study represents the first internal investigation focusing on the role of technostress in the relationship between job satisfaction and individual innovation. Given the increasing proliferation of technology in people's lives (Mohammadi et al., 2023), technostress undeniably stands as one of the most influential factors affecting organizations, institutions, and hospitals. The research results indicate a very strong influence of job satisfaction on individual innovation. Similar findings were reported in the study by Shipton et al., conducted among 3717 employees from 28 British production organizations, indicating that job satisfaction plays a crucial role in innovation within production, organizational, and employee contexts (Shipton & et al,. 2006).

Analyzing the results regarding the relationship between job satisfaction and the components of individual innovation, the evidence suggests that job satisfaction significantly and meaningfully affects idea generation, idea promotion, and idea implementation. Thus, one of the crucial factors affecting individual innovation and its components is job satisfaction. Organizations must consistently work towards increasing employee job satisfaction to enhance individual innovation. To encourage and foster job satisfaction, organizations should provide appropriate rewards to employees who seek new methods at work and assist their colleagues. Managers should support the implementation and easy execution of new ideas. To increase job satisfaction, employees' opinions should be sought before the adoption of new systems, and their assistance should be enlisted in aligning new systems with organizational needs and goals. Intra-organizational systems should be tailored to employees' needs, and an interactive and open environment for communication with managers should be established.

Furthermore, the research results indicate that technostress moderates the relationship between job satisfaction and individual innovation. In examining the components of individual innovation, it was found that technostress moderates the relationship between job satisfaction and idea generation and idea promotion. Therefore, organizations aiming to enhance both individual innovation and job satisfaction must work towards reducing technostress in their environment. Managers should encourage employees to share their knowledge about new technologies and motivate them for teamwork and innovative problem-solving through group work and collaboration. Organizations should limit unnecessary information and emails to control and decrease technostress. Efficient information systems should be implemented and managed within the organization, utilizing knowledge-based companies for organizing and managing internal information. Continuous technical support and assistance should be provided to employees for using new technologies, especially during crises and significant challenges. However, it is crucial to note that technical support requires specific features, such as being accessible to employees, being ready to respond during working hours, and involving knowledgeable and experienced individuals. In conclusion, organizations must prioritize both job satisfaction and the reduction of technostress to enhance individual innovation. These efforts will not only improve employees' well-being but also foster a more innovative and productive work environment.

## 5. Conclusion

In the dynamic realm of contemporary business, understanding the dynamics of innovation is paramount. This study illuminates the pivotal relationship between job satisfaction and individual innovation, elucidating their integral connection. The research reveals a robust and positive correlation between job satisfaction and various facets of innovation, including idea generation and promotion, highlighting the indispensable role content employees play in fostering creative initiatives. Moreover, the study nuances this relationship by introducing the moderating element of technostress. While job satisfaction fuels innovation, technostress, especially in the stages of idea generation and promotion, acts as a significant hurdle. Recognizing this, organizations must prioritize strategies that not only enhance job satisfaction through conducive work environments and recognition but also tackle technostress head-on. In response to these findings, organizations are urged to adopt targeted measures. Elevating job satisfaction can be achieved through employee-centric policies, skill development opportunities, and recognition programs. Simultaneously, managing technostress necessitates investment in digital wellness initiatives and training programs to equip employees with coping mechanisms. In this era of technological interconnectivity, the dual approach of enhancing job satisfaction and mitigating technostress emerges as the cornerstone for fostering workplace innovation. Organizations embracing these strategies are poised not only to nurture a culture of creativity but also to empower their workforce to navigate the challenges of the modern, technology-driven workplace effectively. These proactive steps are vital for organizations aspiring to remain agile, competitive, and innovative, ensuring a resilient future in the face of evolving industry demands.